\documentclass[sigconf,nonacm]{acmart}

\AtBeginDocument{%
  }

\setcopyright{none}
\renewcommand\footnotetextcopyrightpermission[1]{}
\copyrightyear{2027}
\acmYear{2027}
\acmConference[SIGCSE TS 2027]{Proceedings of the 58th ACM Technical Symposium on Computer Science Education}{February 17--20, 2027}{Sacramento, CA, USA}

\usepackage{booktabs}
\usepackage{xtab}
\usepackage{listings} 
\usepackage{enumitem}
\newsavebox{\qbox}
\begin{document}

\title{Can LLMs Simulate Novice Programmers' Misconceptions?}


\author{Batuhan Yeltekin}
\email{batuhan.yeltekin@columbia.edu}
\affiliation{%
  \institution{Columbia University}
  \city{New York}
  \state{NY}
  \country{USA}
}

\author{Daniel Bauer}
\email{bauer@cs.columbia.edu}
\affiliation{%
  \institution{Columbia University}
  \city{New York}
  \state{NY}
  \country{USA}
}

\renewcommand{\shortauthors}{Yeltekin and Bauer}

\begin{abstract}
We evaluate 13 LLMs on generating, solving, simulating, and diagnosing novice programming misconceptions on code-tracing problems. While frontier models reliably perform these tasks, small models ($\le$14B) struggle with dynamic execution. Notably, code-tuned models fail during misconception simulation by reverting to correct execution. Finally, misconception diagnosis is significantly easier in True/False formats than in open-ended generation tasks.
\end{abstract}




\maketitle

\section{Problem and Motivation}
\begin{figure}[t]
\centering
\begin{lrbox}{\qbox}
\begin{minipage}{0.92\linewidth}
\footnotesize
\textbf{Question 34.} What is the exact output of the following program?
\begin{lstlisting}[language=Python,basicstyle=\ttfamily\small,aboveskip=4pt,belowskip=4pt]
lst = []
if len(lst) > 0 and lst[0] == 1:
    print("first element is 1")
else:
    print("safe")
\end{lstlisting}

A. \texttt{safe} ~ B. An \texttt{IndexError} from evaluating \texttt{lst[0]} on an empty list. \\ C. \texttt{first element is 1} ~ D. A \texttt{TypeError}
%
\end{minipage}
\end{lrbox}
\fbox{\usebox{\qbox}}
\caption{MC question for \textsc{NoShortCircuit} (misconception: \texttt{and}/\texttt{or} always evaluate both operands). The correct answer is A. Option B is the \emph{misconception answer}.}
\label{fig:example}
\end{figure}
Large Language Models (LLMs) are gaining importance in CS education, not only as tools for AI assisted software engineering and code generation, but also as systems that assist instructors, such as autograders, or that provide feedback through code review assistants \citep{kiesler2023, azaiz2024}, TA-like chatbots \citep{liu2024}, and tutoring systems \citep{phung2023}. Evaluations of LLMs for code generation emphasize whether they understand specifications, produce correct solutions, and find/fix errors. However, in education, rather than simply detecting and fixing errors, systems  should be able to understand \textit{why} novice programmers make mistakes in their code \citep{kaczmarczyk2010}.
Beginning programmers frequently have incorrect mental models of programs and program execution (the {\it notional machine} \citep{sorva2013}). Skilled instructors can diagnose such underlying misconceptions from student work. In this paper, we focus on code tracing tasks, where students need to mentally simulate execution of a program and predict the output (a common type of exam question). Incorrect predictions may be diagnostic of specific misconceptions. 
We study how well different LLMs simulate this process from four angles.
\textbf{RQ1:} Can LLMs generate new diagnostic code tracing tasks that could be used in quizzes, exams and tutoring systems (given definitions of misconceptions)?; \textbf{RQ2:} Can LLMs solve simple code tracing tasks, common in a CS1 course?; \textbf{RQ3:} Can LLMs be prompted to solve code tracing tasks incorrectly in ways that simulate specific student misconceptions? (i.e. the student perspective)?; \textbf{RQ4:} Can LLMs identify student misconceptions from incorrect answers to code tracing tasks? (i.e. the instructor's perspective)?


\section{Background and Related Work}
The idea that mistakes novices make in programming are not random, but caused by specific and systematic misconceptions, has a long history in CS education \citep{Soloway1986,kaczmarczyk2010}.
 Much work has focused on assembling misconception catalogs \citep{chiodini2021}, through close observation and interviews \citep{duboulay1986} or mining common errors from repositories of student code \citep{altadmri2015}, including with LLMs \citep{al-hossami2026}.
Recent work in STEM education has explored LLMs for misconception modeling. \citet{sonkar2024,zengaffinen2026} finetune models to enable them to model specific student misconceptions about algebra. \citet{liu2025} find that LLM error patterns on math problems do not align with the errors students make. \citet{parker2026} use LLMs to aid instructors in identifying misconceptions in biomedical science. \citet{scarlatos2026} use LLMs to simulate students to testing educational technology.
We are the first to specifically investigate LLMs for misconception modeling in CS education, with implications for teaching and learning programming with generative AI. 

\section{Approach and Uniqueness}

ProgMiscon\footnote{\url{https://progmiscon.org/}} \cite{chiodini2021} is a curated online repository of common misconceptions encountered by students. We downloaded labels and brief descriptions of the 30 available Python misconceptions. For each misconception, we prompted Anthropic's {\ttfamily Claude Fable 5} model to generate two exam-style code tracing problems, one multiple-choice version with four total options and one True/False question, for a total of 60 questions (see example in figure \ref{fig:example}), with additional manual review. One of the answer choices reflected the correct output, while one (the {\it misconception answer}) corresponded to the specific misconception. Every MC question had one correct option and at least one misconception option, and true/false question had a wrong answer corresponding to the misconception. After manual vetting, four questions had to be revised to align better with the description of the misconception (RQ1).
We then prompted 13 LLMs to complete the following tasks for each code tracing question: \textbf{Exp. 1:} Select a response from the provided candidate answers (RQ2); \textbf{Exp. 2:} Predict the output and a brief explanation, without access to candidate answers (RQ2).\footnote{Only 27 out of the 60 generated questions were solvable without access to the answer choices because the answer choices often contain part of the problem statement.}; \textbf{Exp. 3:} Select a response from the provided candidate answers under the pretense of suffering from the misconception (RQ3); \textbf{Exp. 4:} Given a question and {\it misconception answer}, diagnose the misconception (free form output) (RQ4); \textbf{Exp. 5: }  Repeat Exp. 4, but select the misconception from a list of all 30 misconception definitions (RQ4). For Exp. 2 and 4, we matched the LLMs' output against the reference answer or misconception definition, using {\ttfamily Claude Fable 5} with human review.



\section{Results and Contributions}

\begin{table}[!t]
\centering\small
\setlength{\tabcolsep}{4pt}
\begin{tabular}{lrrrr}
\toprule
 & \multicolumn{2}{c}{MC/TF ($n=60$)} & \multicolumn{2}{c}{Open-ended ($n=27$)} \\
\cmidrule(lr){2-3}\cmidrule(lr){4-5}
Model & Correct & Acc. & Correct & Acc. \\
\midrule
Opus 5            & 60 & 100.0\% & 27 & 100.0\% \\
Sonnet 5          & 60 & 100.0\% & 26 &  96.3\% \\
Haiku 4.5         & 60 & 100.0\% & 26 &  96.3\% \\
GPT-5.6 Sol       & 60 & 100.0\% & 25 &  92.6\% \\
GPT-5.6 Terra     & 60 & 100.0\% & 26 &  96.3\% \\
GPT-5.6 Luna      & 60 & 100.0\% & 27 & 100.0\% \\
OLMo-3:7b    & 57 &  95.0\% & 26 &  96.3\% \\
Qwen3-coder:30b   & 58 &  96.7\% & 24 &  88.9\% \\
Qwen3:30b         & 60 & 100.0\% & 25 &  92.6\% \\
Qwen2.5-coder:14b & 58 &  96.7\% & 24 &  88.9\% \\
Qwen2.5:14b       & 57 &  95.0\% & 22 &  81.5\% \\
Qwen2.5-coder:7b  & 50 &  83.3\% & 20 &  74.1\% \\
Qwen2.5:7b        & 50 &  83.3\% & 17 &  63.0\% \\
\bottomrule
\end{tabular}
\caption{Accuracy of solving code tracing problems correctly for MC/TF items (Exp. 1) and open-ended items (Exp. 2).}
\label{tab:solve}

\end{table}

Table \ref{tab:solve} shows results for Exp. 1 and 2.
Large frontier models ({\ttfamily Opus, Sonnet, Haiku} and  {\ttfamily GPT-5.6}) solve the entire question bank with minimal mistakes on the open-ended questions. 
The 30b open-weight models are also competitive, with {\ttfamily Qwen3}'s 30 billion parameter non-coder variant returning a perfect score on MC/TF.\footnote{We used API/CLI access for Anthropic and OpenAI models, and deployed the open weight models, including OLMo~3~\cite{olmo3}, locally using Ollama~\cite{ollama}.} Smaller models appear unable to model dynamic code execution at the same level as students. 
We did not find systematic differences between models fine-tuned on coding tasks and comparable non-coding models.
\begin{table}[!t]
\small\centering
\setlength{\tabcolsep}{4pt}
\begin{tabular}{lrrr}
\toprule
  Model & Followed & Other wrong & Correct \\
\midrule
Opus 5            & 96.7\% &  0.0\% &  3.3\% \\
Sonnet 5          & 90.0\% &  1.7\% &  8.3\% \\
Haiku 4.5         & 85.0\% &  0.0\% & 15.0\% \\
GPT-5.6 Sol       & 95.0\% &  1.7\% &  3.3\% \\
GPT-5.6 Terra     & 86.7\% &  3.3\% & 10.0\% \\
GPT-5.6 Luna      & 88.4\% &  3.3\% &  8.3\% \\
OLMo-3:7b         & 76.7\% &  3.3\% & 20.0\% \\
Qwen3-coder:30b   & 45.0\% &  1.7\% & 53.3\% \\
Qwen3:30b         & 86.7\% &  1.7\% & 11.6\% \\
Qwen2.5-coder:14b & 55.0\% &  8.3\% & 36.7\% \\
Qwen2.5:14b       & 61.7\% &  3.3\% & 35.0\% \\
Qwen2.5-coder:7b  & 53.4\% & 13.3\% & 33.3\% \\
Qwen2.5:7b        & 38.3\% &  6.7\% & 55.0\% \\
\bottomrule
\end{tabular}
\caption{Exp. 3 results. \emph{Followed}: Accuracy of selecting the  {\it misconception answer}. \emph{Other wrong }:problem solved incorrectly by selecting an answer other than a {\it misconception answer}.  \emph{Correct}: misconception was not modeled}
\label{tab:roleplay}
\end{table}
Table \ref{tab:roleplay} shows results for Exp. 3., investigating if LLMs can internalize programming misconceptions. The \emph{Followed} column shows the percentage of MC/TF questions for which the model selected the {\it misconception answer}, i.e. the answer we would expect a student suffering from the misconception would give. Larger models with higher reasoning depth ({\ttfamily Opus} and {\ttfamily GPT-5.6 Sol}), perform best. The \emph{Correct} column shows the percentage of questions for which the model chose the expected answer ignoring the misconception. This was an issue for coder models, which have been trained on large amounts of correct code examples. The misconception description in the prompt is not sufficient to overcome their bias for correct code generation and debugging. Inspecting the reasoning chain for {\ttfamily Qwen3-coder}, we find that it interrupts the process midway and turns from assuming the misconception to attempting to just solve the question. Table \ref{tab:diagnosis} shows results of detecting the misconception from a question / answer pair, either asking for a free description (Exp. 4) or selecting from the list of all misconceptions (Exp. 5). Selecting from a list is somewhat easier and, as before, large reasoning models perform best. Detecting misconceptions from TF question seems easier across both modalities. The TF questions are designed to more concretely address one specific misconception, while MC questions are more open-ended and require more abstract reasoning to identify the misconception from a single response. 

Overall, our results suggest that (RQ1) LLMs can reliably generate diagnostic code tracing problems from misconception definitions; (RQ2) Larger LLMs (30+B parameters) can reliably solve code tracing problems; (RQ3) Large LLMs can mimic student behavior when answering code tracing problems under specific misconceptions. Model size and reasoning abilities seem to be key for this task and coding oriented models appear to be overfitted to a correct (misconception free) understanding of code. Future work will investigate to what degree LLMs actually {\it model} the notional machine and dynamic code execution rather than just stochastically {\it simulating} behavior.

\begin{table}[!t]
\small\centering
\setlength{\tabcolsep}{3pt}
\begin{tabular}{lrrrrrrr}
\toprule
 & \multicolumn{4}{c}{Free (Exp.~4)} & \multicolumn{3}{c}{List (Exp.~5)} \\
\cmidrule(lr){2-5}\cmidrule(lr){6-8}
Model & Acc. & MC & TF & Sim. & Acc. & MC & TF \\
\midrule
Opus 5            & 86.7\% & 76.7\% & 96.7\% & 4.40 & 96.7\% & 96.7\% &  96.7\% \\
Sonnet 5          & 88.3\% & 80.0\% & 96.7\% & 4.42 & 95.0\% & 93.3\% &  96.7\% \\
Haiku 4.5         & 86.7\% & 76.7\% & 96.7\% & 4.32 & 91.7\% & 86.7\% &  96.7\% \\
GPT-5.6 Sol       & 91.7\% & 86.7\% & 96.7\% & 4.50 & 98.3\% & 96.7\% & 100.0\% \\
GPT-5.6 Terra     & 90.0\% & 83.3\% & 96.7\% & 4.40 & 96.7\% & 96.7\% &  96.7\% \\
GPT-5.6 Luna      & 85.0\% & 76.7\% & 93.3\% & 4.37 & 91.7\% & 86.7\% &  96.7\% \\
OLMo-3:7b         & 68.3\% & 56.7\% & 80.0\% & 3.42 & 75.0\% & 63.3\% &  86.7\% \\
Qwen3-coder:30b   & 68.3\% & 50.0\% & 86.7\% & 3.67 & 70.0\% & 56.7\% &  83.3\% \\
Qwen3:30b         & 76.7\% & 66.7\% & 86.7\% & 4.12 & 88.3\% & 83.3\% &  93.3\% \\
Qwen2.5-coder:14b & 73.3\% & 63.3\% & 83.3\% & 3.72 & 70.0\% & 56.7\% &  83.3\% \\
Qwen2.5:14b       & 73.3\% & 56.7\% & 90.0\% & 3.65 & 68.3\% & 53.3\% &  83.3\% \\
Qwen2.5-coder:7b  & 65.0\% & 53.3\% & 76.7\% & 3.40 & 61.7\% & 60.0\% &  63.3\% \\
Qwen2.5:7b        & 55.0\% & 30.0\% & 80.0\% & 3.08 & 53.3\% & 50.0\% &  56.7\% \\
\bottomrule
\end{tabular}
\caption{Misconception diagnosis given a question and {\it misconception answer}. \emph{Free} (Exp.~4): the model describes the misconception in its own words. \emph{List} (Exp.~5): the model selects one of the 30 misconceptions.}
\label{tab:diagnosis}
\end{table}

\newpage
\bibliographystyle{ACM-Reference-Format}
\bibliography{bibliography}

\end{document}